\documentclass[12pt]{iopart}
\usepackage{iopams}

\newcommand{\beq}{\begin{equation}}
\newcommand{\eeq}{\end{equation}}
\newcommand{\bea}{\begin{eqnarray}}
\newcommand{\eea} {\end{eqnarray}}

\def\bdelta{{\bar\Delta}}
\def\hil{{\mathcal{H}}}
\def\un\a{{\underline\alpha}}
\def\la{{\langle}}
\def\ra{{\rangle}}
\def\a{{\alpha}}
\def\b{{\beta}}

\def\d{{\delta}}

\def\Tr{{\mathrm{Tr}}}
\def\id{{\mathbf{1}}}

\begin{document}

\title{Spacetime Coarse Grainings in the Decoherent Histories Approach to Quantum Theory}

\author{Petros Wallden}

\address{Raman Research Institute, Theoretical Physics Group \\
Sadashivanagar, Bangalore - 560 080, India} \ead{petros@rri.res.in;
petros.wallden@gmail.com}
\begin{abstract}
We investigate the possibility of assigning consistent probabilities
to sets of histories characterized by whether they enter a
particular subspace of the Hilbert space of a closed system during a
given time interval. In particular we investigate the case that this
subspace is a region of the configuration space. This corresponds to
a particular class of coarse grainings of spacetime regions. We
consider the arrival time problem and the problem of time in
reparametrization invariant theories as for example in canonical
quantum gravity. Decoherence conditions and probabilities for those
application are derived. The resulting decoherence condition does
not depend on the explicit form of the restricted propagator that
was problematic for generalizations such as application in quantum
cosmology. Closely related is the problem of tunnelling time as well
as the quantum Zeno effect. Some interpretational comments conclude,
and we discuss the applicability of this formalism to deal with the
arrival time problem.

\end{abstract}

\pacs{ 03.65.Xp, 03.65.Yz, 04.60.-m, 04.60.Gw, 04.60.Kz}%

 \maketitle

\section{Introduction}

\subsection{Opening remarks}

Questions that involve time in a non-trivial way are not easily
addressed within the context of quantum mechanics. An alternative
formulation of standard quantum theory that is suited to deal with
such questions is the decoherent histories approach. In particular
we will deal with the probability that a system is found in (or
crosses) a spacetime region using the decoherent histories approach
to quantum theory. In the present paper we will focus in the case
that this region has the simple form of a region of the
configuration space extended in the time direction i.e.
$\Delta\times[t,t_0]$ with $\Delta$ being a region of the
configuration space and $[t,t_0]$ a (parameter) time interval. Both
the arrival time problem and the problem of time in
reparametrization invariant theories are closely tied with this
question.

\subsection{The Decoherent histories approach}

We will briefly review the decoherent histories approach in
nonrelativistic quantum theory described by a Schr\"{o}dinger
equation  \cite{GH1,GH2,Gri,Omn1,Omn2,Hal1,Hal2}. This approach is
designed to deal with closed system where there is no separation
between observer and system. The (predictive) statements are made
about histories of the system and it is well suited to deal with
questions that involve time non trivially, i.e. not only about
one-time propositions. The central object in this approach is the
decoherence functional.
\begin{equation}
D(\underline\a,\underline\a ')= \Tr(C_{\underline\a}\rho
C^\dag_{\underline\a '})
\end{equation}
with $C_{\un\a}$, being the class operator that corresponds to a
history $\un\a$. In particular, $C_{\un\a}$ is given by a string of
time-ordered projectors at times $t_1\cdots t_n$,

\begin{equation}\label{heisenberg class operator}
C_\un\a=P_{\a_n}(t_n)\cdots P_{\a_2}(t_2)P_{\a_1}(t_1)
\end{equation}
and $\un\a$ denotes the string $\a_1,\a_2,\cdots\a_n$. The
projection operators are in the Heisenberg picture,

\begin{equation}
P_{\a_k}(t_k)=e^{iH t_k} P_{\a_k} e^{-i H t_k}
\end{equation}
where the projectors form a complete and orthogonal set, i.e.
\begin{equation}
\sum_\a P_\a=\id
\end{equation}
and

\begin{equation}
P_\a P_\b=\d_{\a\b}P_\a
\end{equation}
This  class operator defined in equation (\ref{heisenberg class
operator}) obeys the following

\beq\sum_\un\a C_\un\a=\id\eeq Note that we could have used an
alternative definition of the class operator that in the case of
standard nonrelativistic quantum mechanics lead to the same
probabilities, namely

\beq C_\un\a=P_{\a_n}e^{-iH(t_n-t_{n-1})}P_{\a_{n-1}}\cdots
e^{-iH(t_2-t_1)}P_{\a_1}\eeq which satisfies

\beq\label{sum rule} \sum_\un\a C_\un\a=e^{-iH(t_n-t_1)}\eeq The
distinction of those two is trivial in nonrelativistic quantum
mechanics but not so in reparametrization-invariant theories as was
noted in  \cite{Halliwell:2005nv} and we will use both definitions
in what follows. We may assign probabilities to the possible
histories provided that we have a set of those such that

\beq\mathrm{Re}D(\un\a,\un\a ')=0\eeq
 for $\un\a\neq\un\a '$. While this condition guarantees that the probability rules will be
obeyed (including the additivity of disjoint regions of the sample
space) we will focus on a stronger condition, namely that both the
real and imaginary part of the off-diagonal terms of the decoherence
functional vanish, i.e.

\beq D(\un\a,\un\a ')=0\eeq for $\un\a\neq\un\a '$. This is related
with the existence of generalized  records \cite{GH2} as well as
considerations of composite systems \cite{Diosi}.

Finally, in the case of set of histories obeying the decoherence
condition, we may assign probabilities to histories, taking the
diagonal terms of the decoherence functional

\begin{equation}
p(\a_1,\a_2,\cdots)=D(\un\a,\un\a)=\Tr(C_\un\a\rho\ C^\dag_\un\a)
\end{equation}
To summarize, in order to deal with a question in the decoherent
histories, we require first to construct a class operator that
correspond to this question. Then make sure that the histories
decohere. The latter can be done either by introducing some
interacting environment or by considering special initial states. In
either case, the system has to obey a decoherence condition.
Finally, if the decoherence condition is satisfied, we assign
probabilities to the histories corresponding to the diagonal terms
of the decoherence functional.

\subsection{Arrival time problem}

The first issue that we will consider is the arrival time problem.
This is the question ``\emph{what is the probability of a particle
entering (crossing) a region $\Delta$ of space  at any time between
$t_1$ and $t_2$}''. This question involves time in a non-trivial
way. It is a \emph{spacetime} question. Standard quantum mechanics
has difficulties  with this kind of question, since all the
propositions refer to one particular time. Time has a strange status
in quantum mechanics since there is no self-adjoint operator
representing it and appears like a parameter
\cite{Hartle:1988kt,Hartle:1988zs,Hartle:1991qg} . Closely related
to this are also questions of tunnelling time (how much time the
particle spends in the forbidden region) as well as in
reparamterization invariant theories, where there is no time (see
below). The literature on propositions in quantum mechanics that
refer to more than one time is long \cite{Wigner}. To deal with
this, people have tried to define time in quantum mechanics as an
operator \cite{Grot,Kholevo,Peres}, or using physical internal
clocks \cite{Hartle:1988kt,Hartle:1988zs} or using path integral
approaches \cite{Hartle:1991qg,Fertig,YaT,Kumar}. In this paper we
will use the decoherent histories approach as in \cite{HaZa} .
Similar considerations were made in  \cite{Charis+Ntina}.

\subsection{Problem of time}

 The questions referring to spacetime regions, as in the arrival
 problem, relates with the so-called ``problem of time''. This
 arises in theories that  have vanishing Hamiltonian. They are
 called reparametrization invariant theories, since anything
 observable is independent of time, in the sense that redefining
 the (parameter) time will leave all the probabilities invariant.
 The most notorious of these theories arises in quantum cosmology,
 where the wavefunction of the universe obeys the Wheeler-De Witt
 equation,

 \begin{equation}
 H\Psi[h_{ij},\phi]=0
 \end{equation}
The wavefunction $\Psi$ depends on the three-metric $h_{ij}$ and the
matter field configurations $\phi$ on a closed three-surface
\cite{Hartle91:Baby,Hartle92:Houches,Hal92}. Other examples are the
parametrized particle and the relativistic particle. In all these
cases, the lack of time parameter forces us to make statements
concerning regions of the configuration space, without singling out
one variable as time. This will be easier to carry out if we have an
effective way to treat spacetime questions in the context of quantum
mechanics as, for example, in the arrival time case.

There are two particular approaches that have made progress in the
quantization of these ``timeless'' theories. The one is the evolving
constants method \cite{Rov1,Rov2,Mar1,Haj,MRT,Mon,GaPo} . The other
that we are going to follow here, is the decoherent histories (for
previous related work see
\cite{HaTh2,HaTh1,Halliwell:2002cg,Halliwell:2005nv}). In particular
the requirements are the following:
\begin{itemize}

\item[(a)] The state $|\psi\ra$ should obey the constraint, \beq
H|\psi\ra=0\eeq

\item[(b)] In the context of a timeless theory, one needs to think
afresh what a class operator is. The lack of time forces us to drop
the simple definition (\ref{heisenberg class operator}). In
particular we require the class operators to commute with the
constraint (Hamiltonian). \beq[C_\a,H]=0\eeq Note also, that the
class operator $C_\a$, is a projector ``classically''. This means
that if all the projectors at different (parameter) time commuted,
the class operator would reduce to a projector.

\item[(c)] The inner product in the decoherence functional should
be the so-called ``induced inner product'' for the states to be
normalized \cite{Rie}. This is needed because the states for most of
these theories, are not normalized in the usual Schr\"{o}dinger
inner product. This is effectively an inner product defined on the
solution surface. See \cite{HaTh2,HaTh1,HaMa} for applications
similar to those considered here.

\end{itemize}
Of course, on top of these, as in all the cases in decoherent
histories, we require decoherence in  order to be able to assign
probabilities. Note that in Section \ref{Section Rep Inv} we will
focus on the case that the initial state is an energy eigenstate,
rather than having vanishing Hamiltonian (which is just a special
case for $E=0$). This would correspond in shifting the total energy
by a constant amount. The class operator will still need to commute
with the Hamiltonian and the only difference is that the constraint
will then be $(H-E)|\psi\ra=0$. It will help us seeing some easy
non-trivial examples, since for most simple models the zero energy
eigenstate is trivial. The latter is clearly not true for the
cosmology case that was mentioned above.

\subsection{This paper}

This paper examines the decoherent histories approach when applied
the arrival time problem and the problem of time in
reparametrization invariant theories. These correspond to a
particular class of spacetime questions. In particular in Section
\ref{Restricted Propagator Section} we will review different forms
of the restricted propagator that is of use, and derive a new
expression. In Section \ref{Section Arrival Time} we will consider
the arrival time problem. First in Section \ref{Section Arrival
condition} we will get a general decoherence condition and we will
explore it in detail in Section \ref{Section arrival details},
analyze some examples in Section \ref{Section Examples}, and discuss
the consequences for the probabilities in Section \ref{Section
discussion}. In Section \ref{Section Rep Inv} we will deal with the
problem of time reparametrization invariant theories and provide a
general decoherence condition. This condition is
 independent of the detailed calculation of the restricted
 propagator and depends solely on the energy eigenfunctions and the
 boundary of the region of interest. In Section \ref{Section rep inv detail} we will also suggest what we can calculate
 in the case of quantum cosmology, following the suggested formalism. We summarize and conclude in Section \ref{Conclusions}.

\section{New expression for the restricted propagator}\label{Restricted Propagator Section}

An important mathematical object in all the discussion of the
arrival time in the decoherent histories, as well as in many other
applications, is the \emph{restricted propagator}. This is the
propagator restricted to some particular region $\Delta$ (of the
configuration space) that corresponds to a subspace of the total
Hilbert space denoted by $\hil_\Delta$. In this section we will
provide several different expressions including a new one and prove
their equivalence. The most common form is the path integral one:

\begin{equation}\label{restricted-path integral}
g_r(x,t\mid x_0,t_0)=\int_\Delta \mathcal{D}x \exp(iS[x(t)])=\langle
x| g_r(t,t_0)| x_0\rangle
\end{equation}
The integration is done over paths \emph{that remain in the region
$\Delta$ during the time interval $[t,t_0]$}. The $S[x(t)]$ is as
usual the action. Note that if there is ambiguity about which region
is the propagator restricted to we will add a superscript (say
$g_r^\Delta(t,t_0)$ for example). The operator form of the above is
given by \cite{Halliwell:2005nv,Halliwell:1995jh}:

\begin{equation} \label{restricted operator}
 g_r(t,t_0)=\lim_{\delta
t\rightarrow 0} P e^{-iH(t_n-t_{n-1})}P\cdots P e^{-iH(t_1-t_0)}P
\end{equation}
With $t_n=t$, $\delta t\rightarrow 0$ and $n\rightarrow \infty$
simultaneously keeping $\delta t \times n=(t-t_0)$. $H$ is the
Hamiltonian operator. $P$ is a projection operator on the restricted
region $\Delta$ and in cases that it is not clear we will add a
subscript (say $P_\Delta$). We therefore have

\beq g_r(x,t\mid x_0,t_0)=\la x|g_r(t,t_0)|x_0\ra\eeq Note here that
the expression (\ref{restricted operator}) is the defining one for
cases that the restricted region is not a region of the
configuration space, but some other subspace of the total Hilbert
space $\hil$. The differential equation obeyed by the restricted
propagator is:

\begin{equation} \label{restricted differential}
(i\frac{\partial}{\partial t}-H) g_r(t,t_0)=[P,H] g_r(t,t_0)
\end{equation}
Which is almost the Schr\"{o}dinger equation, differing by the
commutator of the projection to the restricted region with the
Hamiltonian.

We will show that the restricted propagator can also be expressed as

\begin{equation} \label{restricted zeno}
g_r(t,t_0)=P\exp\left(-i(t-t_0)P H P\right)P
\end{equation}
This relation  will turn out to be the most useful in our paper.
Note that $PH P$ is the Hamiltonian projected in the subspace
$\hil_\Delta$. To prove equation (\ref{restricted zeno}) we multiply
 (\ref{restricted differential}) with $P$ we will then get

\beq (i\frac{\partial}{\partial t}-PHP) g_r(t,t_0)=0\eeq using the
fact that $P[H,P]P=0$ and that the propagator has a projection $P$
at the final time. This is Schr\"{o}dinger equation with Hamiltonian
$PHP$. It  is evident that this leads to the full propagator in
$\hil_\Delta$ provided that the operator $PHP$ is self-adjoint in
this subspace \cite{Facchi2}. We will now prove the equivalence of
equation (\ref{restricted zeno}) with  (\ref{restricted operator}).
Taking the limit $\delta t\rightarrow 0$ implies:

\begin{eqnarray}
g_r(t,t_0)&=&\lim_{\delta
t\rightarrow 0} P(1-iH\delta t)P(1-iH\delta t)\cdots(1-iH\delta t)P\\
&=& \lim_{\delta t\rightarrow 0}\{P+(-i\delta t n)PHP+(-i\delta
t)^2\left(\begin{array}{c} n\\2\end{array}\right)PHPHP+\cdots\}\nonumber\\
&=&\lim_{\delta t\rightarrow 0}\{P+(-i\delta t PHP)+(-i\delta t
PHP)^2\left(\begin{array}{c}
n\\2\end{array}\right)+\nonumber\\&&+\cdots(-i\delta t
PHP)^k\left(\begin{array}{c}
n\\k\end{array}\right)+\cdots\}\nonumber
\end{eqnarray}
Noting that in the limit we consider, that $\delta t\rightarrow 0$
and $n\rightarrow \infty$ keeping $\delta t \times n=t-t_0$:

\begin{equation}
(\delta t)^k\left(\begin{array}{c} n\\k\end{array}\right)\rightarrow
\frac{(\delta t n)^k}{k!}
\end{equation}
We have:
\begin{equation}
g_r(t,t_0)=\lim_{\delta t\rightarrow 0}\{P+(-i\delta t PHP
n)+\cdots+\frac{(-i\delta t PHP n)^k}{k!}\cdots\}
\end{equation}
And therefore we get:

\begin{eqnarray}
g_r(t,t_0)&=& P\exp\{-i(t-t_0)PHP\}P\\
&=&P\exp\{-i(t-t_0)HP\}
\end{eqnarray}
 The above expressions will be used in
the following sections. Finally, note that (\ref{restricted zeno})
satisfies trivially  (\ref{restricted differential}).

\section{Arrival Time Problem in the Decoherent Histories}\label{Section Arrival Time}

From here on we will be using the decoherent histories approach to
quantum theory. Let the configuration space $\mathcal{Q}$ be a
(Riemannian) manifold that for simplicity we will assume to be
$\mathbb{R}^n$ and consider a region $\Delta$ in that. The question
that we will be asking is ``\emph{What is the probability that the
system crosses the region $\Delta$ within the time interval
$[t,t_0]$}''. The most natural way to deal with this is to ask the
probability that the system remains during this time interval, in
the complimentary region $\bdelta$ and deduce the answer to our
question by classical logic (the negation is the answer to the
arrival time question). To apply classical logic we require the
histories to decohere. We will proceed to find what condition the
initial state and the time interval should satisfy for the latter to
be true. Note that the region $\bdelta$ is defined as
$\Delta\cup\bar\Delta=\mathcal{Q}$ and
$\Delta\cap\bar\Delta=\emptyset$.

\subsection{Decoherence condition and probabilities}\label{Section Arrival condition}

The class operator corresponding to the history of remaining in
$\bdelta$ is

\beq C_\bdelta=g^{\bar\Delta}_r(t,t_0)\eeq where the restricted
region of propagation is $\bar\Delta$ corresponding to projector
$\bar P$ (from here on, when referring to restricted propagator, it
will be understood that it is in the region $\bar\Delta$ and we will
thus omit the superscript). The class operator for entering the
region $\Delta$ in that time interval is equal with the crossing
propagator

\beq C_\Delta=g_c(t,t_0)=g(t,t_0)-g_r(t,t_0)\eeq where $g(t,t_0)$ is
the full propagator. The above definition of the crossing propagator
follows directly from the path integral expressions, since the paths
that cross region $\Delta$ are all the paths except those that
remain always in $\bar\Delta$. This class operator, is also
consistent with (\ref{sum rule}). We will assume for simplicity pure
initial state $|\psi\ra$. If the restricted Hamiltonian $H_r=\bar PH
\bar P$ is self-adjoint operator in $\hil_\bdelta$, and using
(\ref{restricted zeno}) we have

\beq\label{restricted-projection} g^\dagger _r(t,t_0)g_r(t,t_0)=\bar
P\eeq We should point out that $H_r$ is indeed guaranteed to be
self-adjoint in $\hil_\bdelta$ in the following two cases

\begin{itemize}

\item[(i)] The subspace spanned by $\bar P$ is finite dimensional.
This relates to the usual account of the quantum Zeno effect
\cite{Misra,Itano,Wilkinson}.

\item[(ii)] The subspace spanned by $\bar P$ is a region
$\bar\Delta$ in the configuration space and the Hamiltonian is
quadratic in momenta. This has been shown in \cite{Facchi2} (see
also \cite{Wallden}).

\end{itemize}
The problems that we will deal with fall in this second class. To
have decoherence, and thus being able to assign probabilities to
histories, we require that the off-diagonal terms of the decoherence
functional vanish.

\beq D(\Delta,\bdelta)=\la\psi|C^\dagger_\Delta
C_\bdelta|\psi\ra=0\eeq Using (\ref{restricted-projection}) this
leads to the very general condition

\beq\label{general decoherence condition}
\la\psi|g^\dagger(t,t_0)g_r(t,t_0)|\psi\ra=\la\psi|\bar
P|\psi\ra\eeq where $g^\dagger(t,t_0)=e^{iH(t-t_0)}$ is the full
propagator. This is the main result of this section and will be
explored in detail in the next section. For this condition to be
satisfied, we will show that the initial state $|\psi\ra$ has to
vanish on the boundary of the region $\bdelta$ that we will denote
as $\partial\bdelta$, that is

\beq \label{vanish boundary}\la x|\psi\ra=0\quad\forall\quad
x\in\partial\bdelta\eeq Note that while the vanishing boundary for
the initial state is a necessary condition for decoherence, it is by
no means a sufficient one. We will analyze the consequences of
equation (\ref{general decoherence condition}) for the general case
in the next section. Here we will give a heuristic argument that the
state vanishing on the boundary is a necessary condition. The point
is that for the state

\beq \label{psi restricted
evolved}g_r(t,t_0)|\psi\ra=|\psi(t)\ra_r\eeq  to exist, i.e. belongs
in the Hilbert space $\hil$ we require the state $|\psi\ra$ to
vanish on the boundary. To see this note that the Hamiltonian
involves a term with $\partial^2$. The projector on a region
corresponds to a characteristic function $\chi_\bdelta(x)$ defined
as

\beq\chi_{\bar\Delta}(x)=\left\{\begin{array}{ll} 1 & x\in\bar\Delta\\
0& x\in\Delta
 \end{array}\right.\eeq The state $\bar P|\psi\ra$ involved in $|\psi(t)\ra_r$ will be discontinuous on the boundary unless $|\psi\ra$
 obeys (\ref{vanish boundary}). It follows that the state $|\psi(t)\ra_r$ will not be square
 integrable and will not belong to $\hil$ if $|\psi\ra$ does not vanish on the boundary due to the delta function that will appear
 from the derivative of a discontinuous function. For more detail the reader is
referred at the \ref{appendix b.c.}.

Given that we have initial state and time interval such that the
histories $C_\Delta$ and $C_\bdelta$ decohere, we can assign them
probabilities. The probability for no-entering the region $\Delta$
 is

\beq p_r=D(\bdelta,\bdelta)=\la\psi|\bar P|\psi\ra \eeq and the
crossing probability

\beq\label{crossing probability} p_c=1-p_r=\la\psi| P|\psi\ra\eeq
 An important thing to note, is that if we had an
initial state belonging to the subspace $\hil_\bdelta$, i.e.

\beq \bar P|\psi\ra=|\psi\ra\eeq the crossing probability would be
zero. This means that for a state localized outside the region
$\Delta$, the probability that it crosses the region $\Delta$ is
zero. This is less problematic than it may seems in the first sight,
since most states will not decohere and therefore the above
``candidate-probabilities'' would not correspond to real
probabilities. We will return to this point in Section \ref{Section
discussion}.

\subsection{A detailed look at the decoherence condition}\label{Section arrival details}

We will  explore the consequences of (\ref{general decoherence
condition}). The time evolved stated $|\psi(t)\ra$ is

\beq |\psi(t)\ra=\exp (-i\hat H(t-t_0))|\psi\ra=g(t,t_0)|\psi\ra\eeq
and the restricted time-evolved state is given by (\ref{psi
restricted evolved}). The decoherence condition requires that the
overlap of those two states at time $t$ is the same as their overlap
at the initial time $t_0$,

\beq \la\psi(t)|\psi(t)\ra_r=\la\psi|\bar P|\psi\ra\eeq This will in
general depend (usually sensitively) on the time $t$. It is
physically reasonable to require the decoherence to persist in time
and not depend sensitively on the time interval.

The restricted time-evolved state corresponds to considering the
region of restriction with same Hamiltonian and introducing infinite
potential walls on the boundary. The state would then be reflected
on the boundary. The decoherence condition (\ref{general decoherence
condition}) is satisfied in the following four cases. The first
three of those are independent of the time interval. In the next
section we will give one distinct simple example of each of those.

\begin{itemize}

\item[(a)] The state $|\psi\ra$ is an energy eigenstate
\footnote{This will be of great interest in the reparametrization invariant case, where all physical states are necessarily energy eigenfunctions. }.

\beq H|\psi\ra=E|\psi\ra\eeq On top of this it has to vanish on the
boundary (\ref{vanish boundary}). If the latter is true, then the
state $|\psi\ra$ is a restricted energy eigenstate, i.e.

\beq H_r\bar P|\psi\ra=E\bar P|\psi\ra\eeq with the same energy as
the initial state. This will be explored in detail in Section
\ref{Section rep inv detail}. It implies that the restricted
time-evolved state is given by

\beq |\psi(t)\ra_r=e^{-iE(t-t_0)}\bar P|\psi\ra\eeq The full
time-evolved state is of course

\beq |\psi(t)\ra=e^{-iE(t-t_0)}|\psi\ra\eeq and therefore their
overlap satisfies the decoherence condition (\ref{general
decoherence condition}).

\item[(b)] The restricted propagator can be expressed by the use
of method of images. Then there will be a class of states that
decohere. The restricted propagator can be written as

\beq g_r(t,t_0)=\bar P e^{-i\hat H (t-t_0)}\left(\id+\sum_n \a_n
\mathbf{R}_n \right)\bar P\eeq with $\bf{R}_n$ corresponding to
reflection due to each image and $\a_n$ is the weight of the image.
By considering the fact that at $t=t_0$ we should get $\bar P$, we
have the following condition on the images

\beq\label{images at t_0} \bar P \sum_n \a_n \mathbf{R}_n\bar
P=0\eeq
 The requirement we get to have decoherence is

\beq\label{images condition} \left(\id+\sum_n \a_n \mathbf{R}_n
\right)\bar P|\psi\ra=|\psi\ra\eeq In that case we have

\beq |\psi(t)\ra=g(t,t_0)|\psi\ra=e^{-i\hat
H(t-t_0)}\left(\id+\sum_n \a_n \mathbf{R}_n \right) \bar
P|\psi\ra\eeq which implies

\beq
\la\psi|g^\dagger_r(t,t_0)|\psi(t)\ra=\la\psi|g^\dagger_r(t,t_0)g_r(t,t_0)|\psi\ra\eeq
and therefore the condition (\ref{general decoherence condition}) is
satisfied. There are many states obeying  (\ref{images condition})
and we can easily generate them. Take a fiducial state $|\chi\ra$
such that

\beq \bar P|\chi\ra=|\chi\ra\eeq This means that the state is
localized in the region $\bar\Delta$. Note that the fiducial state
vanishes on the boundary due to the continuity of the wavefunction.
We define

\beq \mathbf{R}_n|\chi\ra=|\phi_n\ra\eeq We can see that

\beq \bar P\sum_n\a_n|\phi_n\ra=\bar P\sum_n\a_nR_n\bar
P|\chi\ra=0\eeq This implies that the following state obeys
trivially (\ref{images condition}).

\beq|\psi\ra=|\chi\ra+\sum_n \a_n|\phi_n\ra\eeq It is therefore
possible to construct a state $|\psi\ra$ that decoheres for every
fiducial state $|\chi\ra$ that is for every state localized in the
$\bdelta$ region. This is still a very limited class of initial
states that decohere.

We should point out here that restricted propagator can be written
as a function of the full propagator, using the method of images, if
and only if there exist a set of energy eigenstates, vanishing on
the boundary, that when projected on the region $\Delta$ forms a
dense subset of the subspace $\hil_\Delta$, i.e. span $\hil_\Delta$.
This is equivalent with requiring that the restricted energy
spectrum (aka spectrum of the restricted Hamiltonian $H_r$) is a
subset of the (unrestricted) energy spectrum. The latter is not in
general the case. An extreme example where the two spectrums are
completely different is the following.

Consider a particle in an infinite potential well (box) with width
say $(d-a)$. Let us consider the restricted region $\bar\Delta$
being a subinterval say $[b,c]$ such that $(b-c)$ is not a rational
multiple of $(d-a)$. The (non-trivial)  restricted energy spectrum
of the region is totally  different from the (unrestricted) energy
spectrum of the full potential well.

\item[(c)] The full time-evolution (for this time interval) of the
initial state $|\psi\ra$ does not leave the subspace $\hil_\bdelta$
i.e. remains always in $\bdelta$. In that case it is evident that
the restricted and the full propagator coincide and thus condition
(\ref{general decoherence condition}) is trivially satisfied.

\item[(d)] Recurrences.  The full time-evolved state of the system
happens to return to same overlap with the restricted time-evolved
state as initially. This is in general very sensitive in time, and
thus of no physical significance. There is however a case that the
decoherence, while still time dependent, may persist for a small
time interval. We could have a state $|\psi\ra$ that for small $t$
the full time evolution (unitary) is within the region $\bdelta$ as
in case (c) above, but in a later time leaves the region. If this
state, further in the future, has a recurrence, i.e. the full and
restricted propagator coincide again, the decoherence will persist,
at least for a while, until the full evolution crosses the region
$\Delta$.

\end{itemize}
Before proceeding to examples of the above cases, we should point
out what happens in the physically interesting case that the initial
state is localized in $\bdelta$. The time persistent cases (a) and
(b) do not apply. We may get decoherent histories that persist in
time ONLY in case (c) where the system is not supposed to cross the
region $\Delta$ anyway. This will be discussed in Section
\ref{Section discussion}.

\subsection{Examples}\label{Section Examples}

We will see here four simple examples of physical situations where
there are initial states that decohere. Each of those corresponds to
one of the above mentioned cases.

\begin{itemize}

\item[(a)] Consider a simple harmonic oscillator with Hamiltonian

\beq\label{oscillator}  H=\frac12
\left(x^2-\frac{\partial^2}{\partial x^2}\right)\eeq where we have
taken $\hbar=m=\omega=1$. Let the initial state $|\psi\ra$ be an
energy eigenstate and in particular the $n=2$ corresponding to
energy  $E=5/2$. This is given by

\beq \la x|\psi\ra=A (2 x^2-1)\exp(-\frac{ x^2}2)\eeq with
$A=\frac{\sqrt2}2\left(\frac1\pi\right)^{1/4}$ normalization
constant. We want to ask what is the probability that the system
crosses the region $\Delta$. For a general region $\Delta$ this
history would not decohere. If we choose the region in such a way
that the initial wavefunction vanishes on the boundary of this
region, then according to Section  \ref{Section arrival details} the
histories would decohere. We therefore choose the region $\Delta$ to
be the (closed) interval $[-\sqrt {2}/2,\sqrt 2/2]$. The probability
to cross this region $\Delta$ is then given by

\beq
p_c=|A|^2\int_{\sqrt2/2}^{-\sqrt2/2}(2x^2-1)^2\exp(-x^2)dx\approx
0.2\eeq Note also that in this example the restricted propagator
cannot be written with the method of images, since the spectrum of
the restricted Hamiltonian does not coincide with the full
Hamiltonian spectrum. It just shares (at least) one common
eigenvalue.

\item[(b)] Consider a free particle in two dimensions with
Hamiiltonian

\beq\label{Hamiltonian-wedge} H=\mathbf{\hat{p}}^2\eeq We take the
region $\Delta$ to be outside a wedge of of angle $\beta$ where
$\beta=\pi/ b$ with $b$ being an even integer. The restricted
propagator for the region
$\bar\Delta=\{\mathbf{x}\in\mathbb{R}^2\mid 0\leq\theta\leq\beta\}$
is given by the methods of images \cite{wedges}

\begin{equation}
g_r(t,t_0)=\bar P
e^{-iH(t-t_0)}\sum_{n=0}^{b-1}(\mathbf{R}_n-\mathbf{K}_n)\bar P
\end{equation}
where $\bar P$ is the projection in the wedge, and we introduce the
rotation operators

\begin{equation}
\mathbf{R}_n=\int rdrd\theta| r,2n\beta+\theta\rangle\langle
r,\theta|
\end{equation}
\begin{equation}
\mathbf{K}_n=\int rdrd\theta| r,2n\beta-\theta\rangle\langle
r,\theta|
\end{equation}
Consider now an initial state

\beq\psi(r,\theta)=\sin (b\theta)R(r)\eeq where $R(r)$ is any
function of $r$ such that the $|\psi\ra$ is normalized. It is clear
that this state obeys the decoherence condition (\ref{images
condition}). The crossing probability will then be

\begin{equation}
p_c=\langle\psi| P|\psi\rangle= \frac{\int_{\pi/b}^{2\pi}\sin
^2(b\theta)d\theta}{\int_0^{2\pi}\sin
^2(b\theta)d\theta}=1-\frac\beta{2\pi}
\end{equation}
 as expected.

 \item[(c)] Here we
will consider a bound system, that therefore its full time-evolution
remains within a distance from the centre. In particular we will
consider the Hamiltonian of the first example (\ref{oscillator}).
The initial state though will not be an energy eigenstate but
instead a superposition of two. Let us consider the state

\beq\psi(x)=B_1\exp(-\frac{x^2}2)+B_2(2 x^2-1)\exp(-\frac{
x^2}2)\eeq that is superposition of the state with energy $1/2$ with
the ground state with energy $5/2$ and $B_1=\sqrt{2}/2$, $B_2=1/2$
i.e.  such that the total state is normalized. The two states are
with equal weight. The region will be $\bdelta=[-10,10]$. We can see
that the restricted propagator is approximately the same as the full
propagator, since the system is localized in the region $\bdelta$.
The width of the gaussian is two that is much less than the region
we consider. We can therefore see that at least for small time
intervals the off-diagonal terms (approximately) vanish. The
crossing probability in this case is zero. Note that more
complicated examples of this type, could include some interacting
environment that restricts the state in $\bar\Delta$ for the time
interval in question.

\item[(d)] We will now consider a particle in an infinite
potential well of width $2\pi$. The Hamiltonian is

\beq H=-\frac12 \frac{\partial^2}{\partial x^2}+V(x)\eeq where
\beq V(x)=\left\{\begin{array}{ll}0 &\quad-\pi\leq x\leq\pi\\
\infty & \quad x<-\pi,\quad x>\pi
\end{array} \right.\eeq
The region $\bdelta$ that we will consider is $0\leq x\leq\pi$. The
particle in an infinite potential well is a periodic system. If the
width of a well is $\a$ then the energies are

\beq E_n=\frac12\left(\frac{\pi n}{\a}\right)^2\eeq From this we can
see that  $t=\frac{4\a^2}\pi$  is the period of the system. The
restricted propagator corresponds to a free particle evolution in a
well of width $\a=\pi$ while the full propagator is in a well of
width $\a=2\pi$. Let us consider the case that the time interval is

\beq T=t-t_0=16\pi\eeq We note that given initial state $|\psi\ra$
the restricted evolved state at time $t=16\pi$ is

\beq|\psi(t)\ra_r=\bar P|\psi\ra\eeq while the full evolved state

\beq|\psi(t)\ra=|\psi\ra\eeq From this is evident that the
decoherence condition (\ref{general decoherence condition}) is
satisfied. We should note that this depends crucially on the choice
of the time interval.

\end{itemize}

\subsection{Discussion}\label{Section discussion}

 An important question is what happens if the initial state is
localized in $\bar\Delta$. From (\ref{crossing probability})
 we get that the crossing probability is zero (provided we have decoherence and can assign probabilities).
  This result can be in disagreement with classical intuition. Take for example a
free-particle wavepacket localized in positive $x$-axis, with
negative average momentum and ask the question whether it should
cross the region of negative x-axis. Classical intuition would tell
us that the particle would cross the region of negative x-axis.
According to our analysis, the particle would either remain always
in the positive x-axis or the above histories would not decohere and
we would not be able to make a prediction.

Here we should stress that the results of Section \ref{Section
Arrival condition} are independent of the Hamiltonian and therefore
apply when the Hamiltonian includes an environment. In that case,
the projections at each moment of time are of the form
$P_{sys}\bigotimes\id_{env}$. The addition of environment, does
 modify the ``reduced'' dynamics, but most importantly, it does \emph{not} (in general) produce decoherence for histories characterized by
class operators of the form of equation (\ref{restricted operator}).
This would seem natural e.g.  \cite{HaZa} since the addition of
environment (quantum Brownian motion) is known to bring decoherence
of the different (coarse grained) trajectories (e.g.
\cite{Halliwell:1999xh}). However the particular coarse graining
considered here (spacetime) would \emph{not} allow for decoherence
(and thus a definite answer) even for a system coupled with
environment such as quantum Brownian motion \footnote{An exception
exist for the case where the environment restricts the full
evolution in the region $\bar\Delta$ for all the time interval.}.
This turns out to be consistent with the results of Hartle (e.g.
\cite{Hartle:1991qg}), who noted that some spacetime coarse
grainings are ``too strong'' for decoherence.

Mathematically what stops us from getting non-trivial probabilities
for crossing, is that in the continuous limit of projectors, we have
no leakage of probability outside the restricted region. This is
related with what is called the quantum Zeno effect, in Copenhagen
quantum mechanics \cite{Misra}, where continuous observation
(corresponding to continuous projectors) stop the state of a system
from evolving outside the subspace of observation.

This still leaves us with the problem of how to get histories that
would correspond to the question \emph{``did the system crossed the
region $\Delta$ in this time period?''} and agree with classical
intuition. The resolution to this apparent paradox comes from the
fact that the same classical questions can be interpreted in quantum
mechanics in many different ways. In particular instead of following
the most natural coarse grainings defined in this section, we could
try to pose the arrival time question differently. We could do the
following three things:

\begin{itemize}

\item[(a)]The restricted propagator we considered was defined by
taking some discreet number of times $t_n$, projecting on the
restricted region and then taking the continuum limit. We could
rather than taking the continuum limit, use some discreet (but
frequent enough), number of projections \cite{Hal-private}. In
between the projections there will be probability leaking out the
region. A decoherence mechanism (environment) should also be
introduced. This is what was effectively calculated in Halliwell and
Zafiris \cite{HaZa} although this was not stated explicitly.

Note also, that this modification (apart from being less natural)
may have problems if we want to consider reparametrization invariant
theories (as in Section \ref{Section Rep Inv}). This is because the
probability will usually depend on the discretization of the times
of measuring (number of $t_n$'s) and therefore will be ``cut-off''
dependent and fail to be independent of reparametrizations.

\item[(b)] We could ``soften'' the notion of measurement, and use
POVM's (Positive Operator Valued Measures) instead of projection
operators. These are quasi-projectors and form an overcomplete basis
of the Hilbert space. They correspond to projection operators in an
enlarged Hilbert space, but when restricted to the one of interest
stop being orthogonal. They still span all the space. We expect that
using those will allow for probability leak from the restricted
region and therefore get non-trivial probabilities for crossing as
was considered in Anastopoulos and Savvidou \cite{Charis+Ntina}.

\item[(c)] We could have considered a (``smooth'') spacetime
region that is not of the form $\Delta\times[t,t_0]$ (rectangular).
This kind of coarse grainings were also considered in
\cite{Hartle:1991qg}. In particular in the example of the
wavepacket, we could have chosen the region to ``follow'' the
wavepacket, i.e. to be at each time a spatial interval around the
centre of the wavepacket and with width greater than the width of
the wavepacket at that time. This history would decohere (similarly
to the case (c) of Section \ref{Section arrival details}) and give
probability approximately one of remaining in this spacetime region.
Since the centre of the wavepacket with negative average momentum
moves in the negative x-axis, we can \emph{implicitly} deduce that
the system crossed the negative x-axis by using classical intuition.
Note though, that still in this case we would not be able to get
non-zero crossing the boundary of the region, probability due to the
same mathematical restrictions as before.

\end{itemize}

To conclude, we point out that we do not get any contradiction, and
the fact that we can get an answer from one coarse graining while no
answer (i.e. no decoherence) from another is due to the many
inequivalent ways we can pose the same classical question in quantum
mechanics.

\section{Reparametrization Invariant Theories in the Decoherent
Histories Approach}\label{Section Rep Inv}

We will now move to another related issue, the problem of models
with reparametrization invariance. The main motivation for
considering such systems comes from the quantum cosmology and the
notorious problem of time in quantum gravity. Finding observables in
such theories has been proven a difficult task. As was mentioned in
the introduction, there are several approaches to this problem. The
 one that we follow here is with the use of decoherent histories,
 and in particular, following the proposal by Halliwell and Wallden in  \cite{Halliwell:2005nv} for the construction of
 class operators that obey the (Hamiltonian) constraint. The question that we want to answer is ``\emph{did our system cross the region
  $\Delta$ of the configuration space with no reference in time}''.
  This type  of questions, form a general enough set of observables.
  In the decoherent histories analysis, we first have to construct
  the class operators corresponding to the set of physical questions
  we want to address. The second part, is to make sure that these
  class operators decohere and we can assign probabilities, at least
  for a general enough set of initial states (or regions $\Delta$),
  even if the coupling of environment is needed. This second part was
  left out from \cite{Halliwell:2005nv} and will be the focus of
  this section in the light of the results we have from the arrival
  time case. Note, that the connection with the arrival time problem, is
mathematically evident, because of the use of restricted propagators
in the definition of the class operators in both cases.

\subsection{Decoherence condition and probabilities}

Inspired by classical considerations, Halliwell and Wallden in
\cite{Halliwell:2005nv}  defined class operators that are consistent
with the constraint and correspond to not crossing the region
$\Delta$ with no reference in time. This was based on the
observation that a whole classical trajectory is reparametrization
invariant and the proposed class operator was a continuous infinite
temporal product of Heisenberg picture projections operators. This
is expressed as,

\begin{equation} \label{Rep Inv Class operator}
C_{\bar\Delta}=\prod_{t=-\infty}^\infty \bar
P(t)=\lim_{t\rightarrow\infty , t_0\rightarrow
-\infty}e^{iHt}g_r(t,t_0)e^{-iHt_0}
\end{equation}
The class operator for crossing is

\begin{equation}
C_\Delta=1-C_{\bar\Delta}
\end{equation}
The limits in (\ref{Rep Inv Class operator}) exist when the class
operator acts on certain class of states and some of these cases
were examined in  \cite{Halliwell:2005nv}. It will turn out that for
states that obey the decoherence condition, this limit does exist.
The decoherence condition, requires the off-diagonal term to vanish
and gives

\begin{equation}\label{Rep Inv Condition 1}
\langle\psi| C_{\bar\Delta}|\psi\rangle=\langle\psi|
C_{\bar\Delta}^\dagger C_{\bar\Delta}|\psi\rangle
\end{equation}
for a pure state $|\psi\rangle$ that is solution to the constraint
equation (i.e. energy eigenstate, or in the case of problem of time,
zero energy eigenstates). Given that $|\psi\ra$ is energy eigenstate
with energy $E$ the right hand side of (\ref{Rep Inv Condition 1})
becomes

\beq\lim_{t,t'\rightarrow\infty,t_0,t_0'\rightarrow-\infty}e^{iE
(t_0-t_0')}\la\psi|g^\dagger_r(t,t_0)e^{-iH(t-t')}g_r(t',t_0')|\psi\ra\eeq
 Provided the individual limits exist (i.e. the class
operator is well defined) we can take the $t,t'$ limits
simultaneously (and similarly the $t_0,t_0'$ limits).  Using
(\ref{restricted-projection}) this leads to

\begin{equation}\label{Rep Inv prob. non crossing}
\langle\psi| C_{\bar\Delta}^\dagger
C_{\bar\Delta}|\psi\rangle=\langle\psi| \bar P|\psi\rangle=p_r
\end{equation}
which is also the probability for not crossing, provided that we
have decoherence (for those states). The left hand side of (\ref{Rep
Inv Condition 1}) becomes

\begin{equation}\label{4.5}
\langle\psi| C_{\bar\Delta}|\psi\rangle=\lim_{t\rightarrow\infty ,
t_0\rightarrow -\infty} e^{iE(t-t_0)}\langle\psi|
g_r(t,t_0)|\psi\rangle
\end{equation}
where E is the eigenvalue of the energy eigenstate $|\psi\rangle$.
We also have, using (\ref{restricted zeno}),

\bea \label{Rep Inv restr propagator}\langle\psi|
g_r(t,t_0)|\psi\rangle&=&\la\psi|\bar P|\psi\ra+\sum_{n=1}^\infty
\frac{(-i(t-t_0))^n}{n!}\times\nonumber\\
&&\left\{\la\psi|(\bar PH)^n|\psi\ra+\la\psi|(\bar PH)^{n-1}\bar
P[H,\bar P]|\psi\ra\right\} \eea Here, we shall assume that the
state  of our system obeys the following condition

\begin{equation}\label{Rep Inv Condition 2}
\bar P[H,\bar P]|\psi\rangle=0
\end{equation}
This will turn out to be the decoherence condition, since we will
show that it is a sufficient condition to obey (\ref{Rep Inv
Condition 1}).  With initial state obeying (\ref{Rep Inv Condition
2}), equation (\ref{Rep Inv restr propagator})  becomes

\begin{equation}\label{4.9}
\langle\psi| g_r(t,t_0)|\psi\rangle=\langle\psi| \bar P|\psi\rangle
\exp (-i(t-t_0)E)
\end{equation}
due to the fact that $(\bar PH)^n|\psi\rangle=E^n\bar P|\psi\rangle$
for these states. Using (\ref{4.9})  the limits in the class
operator in (\ref{4.5}) become trivial, since the dependence on $t$
and $t_0$ disappears. It is now clear that the class operator acting
on states that obey this condition is well defined. We therefore get

\begin{equation}\label{Rep Inv Class operator expectation}
\langle\psi| C_{\bar\Delta}|\psi\rangle=\langle\psi| \bar
P|\psi\rangle
\end{equation}
We  can now see, that from equations (\ref{Rep Inv prob. non
crossing}) and (\ref{Rep Inv Class operator expectation}) the
decoherence condition (\ref{Rep Inv Condition 1}) is satisfied. The
assumption made about the initial states was the equation (\ref{Rep
Inv Condition 2}) which turns out to be the necessary and sufficient
condition to have decoherence for histories corresponding to the
class operators of the form (\ref{Rep Inv Class operator}). In the
case of decoherence, we can finally, assign the probability given by
(\ref{Rep Inv prob. non crossing}) for histories that do not cross
while for histories that cross the probability is

\beq p_c=\la\psi|P|\psi\ra\eeq

\subsection{A detailed look on the decoherence
condition}\label{Section rep inv detail}

Let us first, have a closer look on the decoherence condition
(\ref{Rep Inv Condition 2}). It gives

\begin{equation}\label{Rep Inv Condition 3}
\bar PH\bar P(\bar P|\psi\rangle)=E(\bar P|\psi\rangle)
\end{equation}
where $E$ is the energy of the state $|\psi\rangle$. This is the
general result. We require that there exists a state in
$\mathcal{H}_{\bar\Delta}$ that is restricted energy eigenfunction
having the same energy eigenvalue with the total energy of the state
$|\psi\rangle$ and that vanishes on the boundary. Examples of this
sort were discussed in  \cite{Halliwell:2005nv}. Note here that
following  \ref{appendix b.c.} for the condition (\ref{Rep Inv
Condition 3}) to be satisfied we require a state $|\psi\ra$ such
that

\beq \label{rep inv general decoh condition}\la
x|\psi\ra=0\quad\forall\quad x\in\partial\bar\Delta\eeq The
vanishing boundary conditions is a necessary condition for the sate
to obey condition (\ref{Rep Inv Condition 3}) and therefore
decohere. In the reparametrization invariant case, that we discuss
here, we restrict our attention to initial states $|\psi\rangle$
that obey the constraint, i.e. are energy eigenstates with energy
$E$ determined by the constraint. If we restrict our attention to
those states, the condition (\ref{Rep Inv Condition 3}) reduces to
the requirement that the wavefunction $|\psi\rangle$ vanishes on the
boundary, i.e the latter is a sufficient condition as well. Here we
will show the latter, that proves the claim made in Section
\ref{Section arrival details} at the (a) case. We have

\begin{equation}
H|\psi\ra=E|\psi\ra\quad
\end{equation}
therefore
\begin{equation}
\bar PH(\bar P+P)|\psi\ra=E\bar P\psi\ra\quad
\end{equation}
which implies
\begin{equation}
\bar PH\bar P|\psi\ra+\bar
 P[H,P]|\psi\ra=E\bar P|\psi\ra
\end{equation}
This implies that the state $\bar P|\psi\ra$ is a restricted energy
eigenstate with energy $E$, provided that

\beq \bar P[H,P]|\psi\ra= 0 \eeq This is indeed the case if and only
if the state $|\psi\ra$ vanishes on the boundary. For a quadratic in
momentum Hamiltonian, and with $\Delta$ being a region of the
configuration space we get \beq[H,P]\propto\hat p
\d(x-a)+\d(x-a)\hat p\eeq for all $a\in\partial\Delta$. We therefore
have $\bar P[H,P]|\psi\ra= 0$ in the case that the state vanish on
the boundary and thus equation (\ref{Rep Inv Condition 3}) is
satisfied. Another way to see that this is indeed the case is the
following. To find the restricted energy eigenfunctions we need to
solve the Schr\"{o}dinger equation subject to vanishing conditions
on the boundary. A total energy eigenfunction that vanishes on the
boundary, clearly obeys the restricted Schr\"{o}dinger equation and
therefore is a restricted energy eigenstate. Note that the converse
is not true. We could have other wavefunctions being restricted
energy eigenstates and not total energy eigenstates but we restrict
our attention to the total energy eigenstates since these are the
states that obey the constraint.

We can now  say that for the reparametrization invariant theories a
necessary and sufficient condition to have decoherence, i.e. to obey
 (\ref{Rep Inv Condition 3}), is that \emph{the state $|\psi\ra$
has to vanish on the boundary of the region} i.e. satisfy the
condition (\ref{rep inv general decoh condition}). This is the main
result of this section. This condition is very restrictive in
quantum cosmological models. To this end, we should say that the
following two things can be done in order to meet this conditions.

\begin{itemize}

\item[(i)] For a given state-solution of the Wheeler-DeWitt
equation we can ask what is the locus of the zero's of that
wavefunction. From this we can deduce for which regions we can ask
whether or not the universe was in, by considering a closed region.
This is difficult in general since the solutions are not known, but
it can be done for mini-superspace models at least in the
semi-classical (WKB) approximation.

\item[(ii)] We can do the converse. Given a particular region, we
may search for solutions of the Wheeler-DeWitt equation which have
this region as locus of their zero's.

\end{itemize}

\subsection{Arrival Time-Reparametrization relation}

In the arrival time problem, if we had restricted our attention on
energy eigenstates (case (a) in Section  \ref{Section arrival
details}) we would come to the same condition as in the
reparametrization invariant case. This is indeed expected, since in
that case, the decoherence condition as well as the probability of
crossing are both independent of the time interval. Extending the
interval from minus to plus infinity is not going to change
something.

\section{Summary and Conclusions}\label{Conclusions}
 We have discussed the possibility of assigning consistent
probabilities to histories of the system being in a particular
region of the configuration space continuously and complementary to
this, the probability of crossing a region of the configuration
space. Our analysis relies on the decoherent histories approach to
quantum theory. The questions considered, correspond to a particular
class of spacetime coarse grainings. The main focus of the paper was
to get the conditions that the initial state should satisfy (given
dynamics of system and environment) in order to have decoherence in
the above mentioned types of coarse grainings. We first considered
the arrival time problem applying the most natural coarse graining
related. Using the new expression for the restricted propagator
(\ref{restricted zeno}), we derived the general decoherence
condition (\ref{general decoherence condition}). Only in those cases
we may assign consistently probabilities in the decoherent histories
analysis. We investigated the consequences of this condition and
found that very few initial states satisfy it. Moreover, those
histories gave always zero crossing probability. We discussed this
and suggested that the above coarse grainings are ``too strong'' to
allow decoherence in the arrival time questions. This implies that
we will have to deal with this problem alternatively implementing
the idea that the same classical question correspond to many
inequivalent quantum ones. In particular we suggested considering
projecting on this region at discreet time intervals (as opposed to
the continuous that we used), weakening the notion of measurement
and using POVM's instead of projectors and finally using spacetime
coarse grainings that are not of the form $\Delta\times[t,t_0]$.

We then proceeded by considering a closely related issue, that of
reparametrization invariance. The use of the decoherent histories
for dealing with this aspect of the problem of time is widely
considered as promising. In particular we considered the set of
reparametrization invariant class operator defined in
\cite{Halliwell:2005nv} in the light of the results we had for the
arrival time problem. We then concluded to a very simple general
decoherence condition (\ref{Rep Inv Condition 3}) that is
independent of the precise details of the restricted propagator
contrary to what was previously believed. This condition turns out
to be that the initial state, which is an energy eigenfunction in
order to satisfy the constraint, has to vanish on the boundary of
the region of the configuration space in question. This is a
necessary and sufficient condition. This completes, at least in
formal level, the analysis of the proposal of
\cite{Halliwell:2005nv} to deal with the problem of
reparametrization invariance. We also suggested what can be done
using these results in the case of quantum cosmology that is left
for a future work.

\ack The author is very grateful to Jonathan J. Halliwell for many
useful discussions and suggestions and would like to thank Carl M.
Bender and Chris Isham as well.

\appendix

\section{Vanishing Boundary Conditions}\label{appendix b.c.}

\setcounter{section}{1}

Here we will explore what the decoherence conditions (\ref{general
decoherence condition})  and (\ref{Rep Inv Condition 3}) tells us
about the boundary conditions. First we should point out that the
choice of whether the boundary $\partial \Delta$ belongs to $\Delta$
or  $\bdelta$ is irrelevant. That is true because the Hilbert space
$P_{cl}\hil=\hil_{\Delta cl}$ corresponding to projection in a
closed region is the same with $P_{op}\hil=\hil_{\Delta op}$ that
corresponds to an open region. The reason is that the physical
states, members of $\hil$ ($=L^2[\mathbb{R}]$) are defined as
equivalence classes of ``almost everywhere''  equal functions. Note
that two functions $f,g$ are almost everywhere equal, if there are
equal everywhere except to measure zero subsets. We therefore have
in the norm of the Hilbert space $||f-g||=0$.

Consider a state $|\psi\ra$ that does not vanish on the boundary.
The (unormalized) state $P|\psi\ra$ will have divergent energy
expectation value. Strictly speaking the Hamiltonian operator is not
defined on those states since its domain is AC$ ^2[\mathbb{R}]$,
that is the set of functions in $L^2[\mathbb{R}]$ whose weak
derivatives are in AC$[\mathbb{R}]$. (AC$[\mathbb{R}]$ is the set of
absolutely continuous functions whose weak derivatives are in
$L^2[\mathbb{R}]$). It could not therefore be defined on $P|\psi\ra$
if the state does not vanish on the boundary. We could still think
of expressing it as the limit of a sequence of states belonging at
$\hil_\bdelta$ (that vanish on the boundary and form a dense subset)
where the state converges to $\psi(x)$ everywhere except at the
boundary that the sequence vanishes. We can see that states with
more and more terms in this sequence have higher and higher energy
and can claim that in the limit it will eventually have infinite
energy. This was considered as introducing instantaneously  a hard
wall potential at a point that the wavefunction does not vanish,
first by  Berry in  \cite{Berry} as well as by Bender et al in
\cite{Bender}. The evolution of these states was shown to be
unitary, but resulting to fractal wavefunction with infinite energy.

In the arrival time decoherence condition (\ref{general decoherence
condition}) we had the overlap of the full evolved state with the
restricted evolved state. The latter would correspond to introducing
instantaneously an infinite potential and its evolution as was noted
in  \cite{Berry,Bender} leads to a fractal wavefunction. This means
a wavefunction that is zero at most points, but explodes to infinity
at some zero (or small) measure intervals. Clearly the overlap of
any of these states with the full evolved state that would be a
reasonably well behaved state, is close to zero. This means that it
is \emph{not} equal with $\la\psi|\bar P|\psi\ra$ as it has to be in
order to decohere.

Let us now consider the condition (\ref{Rep Inv Condition 3}) for
the reparametrization invariant case.

\beq \bar PH\bar P|\psi\ra= E\bar P|\psi\ra\eeq This corresponds to
an eigenfunction equation in the subspace $\hil_\bdelta$. If the
energy is finite, i.e. $E\neq\infty$ we need to have $|\psi\ra$
vanishing on the boundary, since otherwise we would get infinite
energy. Note that for the derivation of the above we used the fact
that

\beq g^\dagger_r(t,t_0)g_r(t,t_0)=\bar P\eeq which is the case when
acting to states that follow unitary evolution (i.e. reversible). As
it was analyzed in the above mentioned papers \cite{Berry,Bender}
the sates do indeed evolve unitarily.

Let us illustrate the above with a simple example. Let the region
$\bdelta$ be $[0,\pi]$ and $H=\hat{p}^2$.

\bea u_n(x)&=&\sin nx \quad x\in[0,\pi]\nonumber\\
&=& 0\quad x\in(-\infty,0)\cup(\pi,\infty)\eea These are a complete
set of states (a dense subset) of $L^2[0,\pi]$. Let the state
$P|\psi\ra=|f\ra$ and $\la x|f\ra=f(x)$. We can therefore write

\beq f(x)=\sum_{n=1}^\infty f_n u_n(x)\eeq This series will converge
on $f(x)$ if $f(0)=f(\pi)=0$ everywhere. Otherwise it will converge
to $f(x)$ everywhere except at $0$ and $\pi$. The average energy of
this state will go as

\beq \la f|H|f\ra\propto\sum_{n=1}^\infty |f_n|^2 n^2 \eeq This
series will diverge unless $f_n$ vanishes sufficiently quicker than
$n^2$ as $n$ goes to infinity. It can been shown, that for this to
be the case $f(0)$ and $f(\pi)$ should both vanish. In the case they
do not, the steeper the slope (i.e. more terms in the series meaning
closer to the actual function), the bigger the energy of the state.
It can thus be argued, that requiring the introduction of a rigid
wall ``instantaneously'', i.e using exact projections, would lead to
an infinite energy state.

To conclude, we noticed that the states that do not vanish on the
boundary evolve to fractal wavefunctions and therefore their overlap
with the full evolved state does not obey the condition
(\ref{general decoherence condition}). These states also have
infinite (average) energy in the subspace $\hil_\bdelta$ and
therefore fail to satisfy the decoherence condition (\ref{Rep Inv
Condition 3}) as well.


%

\end{document}